# Truxen: A Trusted Computing Enhanced Blockchain

Chao Zhang

Jason.zhang@truxen.org

## ABSTRACT

Truxen is a Trusted Computing enhanced blockchain that uses Proof of Integrity protocol as the consensus. Proof of Integrity protocol is derived from Trusted Computing and associated Remote Attestations, that can be used to vouch a node's identity and integrity to all of the other nodes in the blockchain network. In this paper we describe how Trusted Computing and Proof of Integrity can be used to enhance blockchain in the areas of mining block, executing transaction and smart contract, protecting sensitive data. Truxen presents a Single Execution Model, that can verify and execute transaction and smart contract in a solo node, consequently enables remote calls to off-chain applications and performs in-deterministic tasks.

## 1. INTRODUCTION

Blockchain was introduced with bitcoin which is an implementation of Satoshi's paper *Bitcoin: A peer-to-peer electronic cash system*. Bitcoin is a so-called decentralized ledger, whose data is opened to everyone holding the ledger, and it does not rely on a central system to maintain the account balances and transactions. The decentralized ledger allows users to hold cash by trusting the cryptographic algorithms rather than trusting authorities like banks.

However, blockchain goes beyond the scope of cash, its character of un-shadowing any transactions enable great potential use scenarios, for example, finance, supply chain, legal evidence and digital assets, especially the domains without trust substantially. On the other hand, blockchain is used in the hostile environment in most cases, where adversaries' attacks and malwares occur frequently. So, some methods and algorithms are used to reach consensus, such as mining blocks to preserve transactions, ECDSA signing to prevent forged transactions, executing transaction individually to calculate results without trust anyone, etc. Some of these methods and algorithms are complex and low efficient, which restrict increasing the throughput and protecting the sensitive data.

Trusted Computing can be used within blockchain to reach consensus straightforward, which mitigates the heavy dependency on hash style mining jobs, and reduces redundant transactions execution. In the meantime, Trusted Computing can be used to protect the credentials, keys, and other personal sensitive data in blockchain.

**Our contribution** in this paper are as follows:

- We present a Proof of Integrity protocol with Trusted Computing enhancement for blockchain, that can reduce the consensus complexity. Trusted Computing seals the identity and integrity that cannot be forged, so the mining process will defend the Sybil, and PoW, PoS, DPoS are not required any more.
- We present a Single Execution Model that allows the transactions and smart contracts are executed on one node only, distributed execution is no longer required to resolve blockchain's efficiency problem.
- We present an approach under the Single Execution Model allowing off-chain invocation and in-deterministic computing, that are missing from other blockchains. Off-chain invocation and in-deterministic computing are critical when introducing blockchain into enterprise application.

Truxen uses the features of Trusted Computing to enhance the blockchain efficiently and is described in the following sections. We also provide a reference code base (https://github.com/truxen-org/chainpoc) for proof of concept.

## 2. TRUSTED COMPUTING

Trusted Computing is a set of specifications proposed by TCG, and it's widely used in many scenarios, like platform integrity check, digital rights management, disk encryption, IoT data protection and critical applications restraint. With the enablement of Trusted Computing, a system will behave consistently in expected ways, and its behaviors will be enforced by combination of hardware and software [18]. Trusted Computing employs a hardware module as a secure crypto processor, named Trusted Platform Module (TPM), to accomplish the unique tasks of Trusted Computing protocols. TCG defines four types of TPM: discrete TPM, integrate TPM, firmware TPM and software TPM, and offers different secure level, among which discrete TPM is the first level and is adopted by Truxen. So, in this paper, terms Trusted Platform Module and TPM are regarded as discrete one, but other types can also work.

Trusted Platform Module integrates a set of cryptographic algorithms, seals cryptographic keys, and offers the facilities to Trusted Computing protocols, e.g. measuring the platform integrity, generating integrity reports, signing data and encrypting sensitive information. These processes run inside the TPM chip and are separated from the other parts of a system. Any malicious code tampering to the system does no impact the execution environment within TPM. Detailed protocol specifications are in Trusted Computing Group's documentation [4][17]. Figure 1 is a reference physical architecture of Trusted Platform Module.

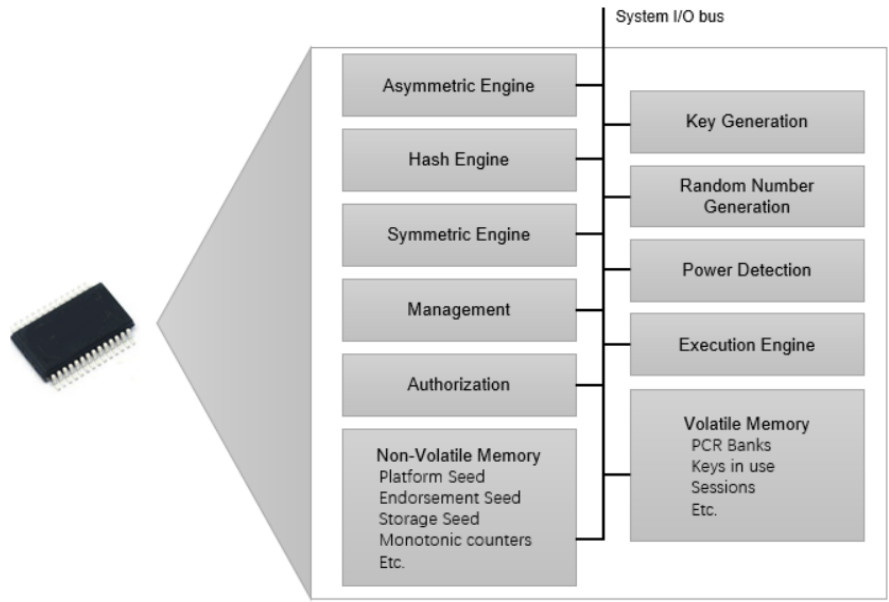
Figure 1: Trusted Platform Module Components

## 2.1 Integrity Measurement

Integrity measurement is the key feature of Trusted Computing. It starts on Root of Trust for Measurement (RTM). The Root of Trust for Measurement is the root of controlled measurement of the integrity of a system, and is followed by 3 levels measurement in the order: static measurement, dynamic measurement and integrity measurement architecture.

When the system is reset, it starts to execute the static integrity measurement before any firmware and application startup, and the static measurement generates integrity values of firmware and bootloader that represents the pre-boot state of the platform. Next, the dynamic measurement takes over the measurement of loading the system kernel. Intel TXT technology is a typical implement of dynamic measurement. Finally, the integrity measurement architecture (IMA) measures the applications, libraries and modules on the top during each of these components starts to run. These integrity values are hashed and extended into several shielded registers in Root of Trust for Storage, called PCRs, which cannot be modified from outside in order to resist the adversaries. In the meantime, PCR event logs are recorded as well. These event log entries contain the PCR number, the value that was extended into the PCR and a log message (giving details what was measured). The messages could be used to track every measurement step and inspect each of the firmware, kernel and applications running on host. Verifier can verify the logs consistency by re-calculating the PCR value from the logs and compare with the original value.

Root of Trust for Reporting can read the PCRs value, and then digitally signs the value with its private attestation signing key inside TPM. The signed PCRs value can be sent out to remote node to vouch for its integrity. The private attestation signing key is encrypted by Root of Trust for Storage key within hardware, which never leave the hardware [31], therefore, nobody can get the key to prevent the signature from forging. However, attestation signing key may also be used to sign any data from outside for general purposes, so the values of internal state consistently start with a special flag TPM_GENERATED_VALUE to ensure they are from inside Trusted Platform Module. Any external value with this flag will be refused to generate signature to prevent forging reporting data.

The integrity measurement flows as following steps:

1) An integrity measurement process retrieves the firmware, kernel and applications packages during starting up, calculates hash values and appends them into TPM PCRs one by one. Any changes to the codes and packages will result in a different measurement value.
2) A reporting process collects the measurement values from PCRs and signs the values with the attestation signing key to generate the Attestation Quote. This process is restricted inside the Trusted Platform Module and nobody can forge the PCRs and signatures.
3) The Attestation Quote and Attestation Key Identity Certification are composed as Integrity Report and sent to remote sites to vouch for its integrity. Attestation Key Identity Certification is a type of certificate and is issued by a third party such as an attestation Certificate Authority (CA), in order to prove the specific attestation signing key is linked with it. Attestation Key Identity Certification is public to all others, and is usually used to verify the signature in Attestation Quote from the TPM.

The Integrity Measurement runs on the local system, and the core part run within TPM. The Integrity Report is measurement result, with which the remote site is able to determine if it's running on the predefined way through Remote Attestation.

## 2.2 Remote Attestation

Remote site who will cooperate with the vouching site needs to evaluate the Integrity Report and determine an appropriate response or action. Remote site validates Attestation Key Identity Certification first to check if it is issued by a trusted third party and linked with a genuine TPM. In a well-organized consortium, Attestation Key Identity Certification is issued if and only if its attestation signing key truly resides in the genuine TPM. A TPM Endorsement Key (EK) helps to provide the residency evidence [4]. Secondly, remote site verifies the Attestation Quote to check if the integrity value (PCR) is correct and the signature is really signed by this TPM. If all of these are inerrant, the system is regard as honest and not tampered.

Attestation Key Identity Certification can be generated and used under a so-called Direct Anonymous Attestation (DAA) architecture which is proposed in TCG specifications. The certification is randomized by the TPM and the host platform before they are sent to the remote sites in order to keep the privacy of the identity. It provides two modes of signatures, anonymous and pseudonymous. The former means different signatures from the same signer cannot be correlated, and the latter means different signatures from the same signer can be correlated but the identity of the signer is still unknown. Detailed information and algorithm can be found in [4][35][38].

It depends on the business requirement to use DAA or not. If the privacy is not a concern, e.g. in Truxen, a simple attestation is straightforward and good enough, otherwise DAA architecture need be setup and followed.

## 3. BLOCKCHAIN ENHANCEMENT

Trusted Computing has strong capability to resist the malicious codes and attacks against adversaries, consequently, all codes, data and functions in such a platform can be trustworthy and consistent. In this section, we leverage Trusted Computing to enhance blockchain in order to reduce the blockchain consensus complexity, and increase the performance, efficiency and security. Finally, we present an innovative Single Execution Model for blockchain.

The mining process implemented with Trusted Computing reduces the complexity of consensus work, because all miner nodes can provide evidence (Integrity Report) that they are enforced to be honest and never submit an invalid block or collude to tamper the historical transactions and balances. The transactions executed by a miner are sufficiently trusted to all the other nodes, no more executions are required, as a consequence, the load in a node is decreased. The result of smart contract calculated by a miner is safe to be persisted into local chain storage in each node. The random number generator with an entropy source inside the TPM can create the true randomness to be used for in-determined purpose, e.g. seed generation and sortition. The sensitive data, like private keys, are encrypted with the storage key by TPM, then stored securely in the system storage. We discuss these details from the next section.

### 3.1 Enhance Mining Process

PoW mining algorithm is widely adopted since the beginning of bitcoin, and demonstrate its security ability in the wild during 9 years. It's simple and sophisticated, which almost 80% of the cryptocurrencies are utilizing as mining consensus. PoW is low efficient as it consumes a large number of electricity to do the hash calculating on each of miners.

However, PoW is a good approach to resist Sybil attack, that is a major issue in the decentralized environment, as public blockchain is totally open to anyone in the internet without any membership control. The only way attackers could launch effective Sybil attacks is to occupy a big number of mining machines, that would cost thousands of millions of money.

The alternative mining algorithm are PoS, DPos, Byzantine Fault Tolerance, PoA, PoP [32], etc., but none of them has the ability to balance the performance, decentralization, efficiency, fairness and operability. There are a lot of research articles on these terms, and detailed discussion is beyond the scope of this paper.

In this section, we propose a Trusted Computing based mining. The mining node on such a base can be regarded as honest and the consensus can be reached concisely, therefore the complexity of mining would be reduced as following reasons:

1) Honest node works as design consistently, nobody can tamper the behavior and deviate the consensus algorithms. Any unpermitted change on the mining node will result in the failure of remote attestation, subsequently the blocks produced by this node will be ignored and eliminated by others from the blockchain network.
2) The miner electing process can be either round-robin or randomized approaches. It is realizable because every Trusted Computing enhanced node have a unique identity which cannot be forged to resist Sybil, comparing to PoA whose identity is stored in a file system that is forgeable and less secure. The round-robin or randomized is quite fair to each miner, no matter how many CPU cores, memories, storages and bandwidths it has. On the other hand, the algorithm can be tweaked to have some weight factor, e.g. some specific miners have more opportunity to produce blocks if they are living for a longer time. This electing process does not act as the competitiveness approach like PoW/PoS, and mitigates the fork and uncle issues on chain. The fork and uncle issues lead to the on-chain transactional data inconsistent as the time goes on until a certain number of blocks are appended after these data.
3) The new block received by a node could be appended to the local chain without executing the transactions, if and when the remote attestation is passed. This can decrease computing effort in a node when blocks arrive at a high frequency (initial downloading bulky blocks is at such a high frequency when a new blockchain client starts, CPU rate is up to around 100% as it executes transactions intensively), so it increases the performance in each node. More details are illustrated in section *Enhance the Performance of Blockchain.*
4) Majority attack is impossible. 51% hash rate would revise the bitcoin transactions history to cause the double spending problem, but it's impossible in Trusted Computing approach because there is no way to change the behavior of the nodes protected by Trusted Computing.

The mining algorithm is very flexible, and most security problems are resolved by Trusted Computing, decreasing the dependent on software cryptographic algorithm. Intel created a PoET [33] mining protocol in the Sawtooth project in conjunction with Intel SGX capability, which is used to elect the miner by waiting a verifiable random time, shortest waiting time is the winner. Microsoft released a whitepaper on a Coco Framework [34], which enables high-scale, confidential blockchain networks using Intel's SGX and Windows Virtual Secure Mode (VSM). In this paper we propose a novel protocol: Proof of Integrity.

## 3.2 Proof of Integrity Protocol

Proof of Integrity Protocol is our major contribution in this paper, it has several processes to reach a consensus among all of the blockchain participants. The *miner join* process checks if a new miner is eligible to join the blockchain and mine. The miner must show their identity and integrity, then others will check if the certificate of identity is issued and signed by an authority, and the integrity value is within a predefined list. The *miner electing* process can be either round-robin or randomized, one miner is elected at a specific round according to the electing algorithm to produce a block.

### 3.2.1 Notions

**Miner** is a node with Trusted Computing enhancement to produce blocks in blockchain network. In most cases, Trusted Computing refers as a computer with TPM. Miner has an identity sealed inside the TPM, and a public certificate to show others that it truly has this identity. The identity can be the attestation signing key or the endorsement key, and the certificate must be issued and signed by an authority (issuer), e.g. the CA. The miner can provide evidence of its system integrity (Integrity Report) to remote nodes to vouch that it's running consistently and honestly without any modification and tampering.

**Miner List** is a list containing all of the valid miners' labels and certificates. A new miner will be added into this list when it passes the *miner join* process. The miner list is located on each of the blockchain node, and it is used to decide which miner is responsible to produce a block at the specific round.

**Integrity** is a value of system integrity through measurement. The measurement happens when system starts up, include the hardware, firmware, OS kernel, module and applications. The Integrity values are persisted in PCRs inside the TPM module, nobody can modify or forge them. The *quote* command can retrieve them with an identity signature.

**Integrity List** is a list containing all of the valid integrity values, and is located on each of the blockchain nodes. The miner whose integrity value falls in the list should be regard as honest.

### 3.2.2 Miner Join Method

When a new miner joins the blockchain network, other nodes will check its qualification and determine the acceptance. The miner join method is as following steps.

1) A new miner sends a join request to blockchain network. The request contains an Integrity Report which is consist of identity certificate and Attestation Quote (quoted integrity values).
2) The active miner for the next round adds the request into a block. So, other nodes can retrieve the request as soon as they get the block, and if a node starts from scratch, it can restore all joined miners after traversing the blocks from the genesis to the latest.
3) Other nodes who get the request from the block, verify the requester's identity

certificate to determine if it is issued by an authorized issuer. This requires the nodes have the issuers' metadata, for example, CA's root cert or intermedia cert.
4) Other nodes verify the Attestation Quote. This requires the nodes to check the integrity values and the signature. The integrity values must fall in the integrity list, and the signature must be signed by requester's identity that can be verified with identity certificate.
5) Other nodes add this miner into its miner list.

Step 3 and 4 are strong constraint in miner join process, as the malicious miner cannot forge the identity, identity certificate and Attestation Quote. The identity (attestation signing key) is encrypted and sealed inside the TPM, and there is no way to reveal and copy the key. The identity certificate is derived from the identity and signed by the authority. Attestation Quote proves the integrity value is truly from the specific TPM without any modification.

### 3.2.3 Miner Electing Method

The principle of miner electing is controlled by the blockchain consensus, which directs the nodes whether or not they should accept a miner's block at a certain round. The original consensus mechanism by Satoshi Nakamoto is one CPU one vote, and every CPU has the same probability to produce blocks. So, the PoW was born when Bitcoin was created. Substantially, Satoshi's consensus is to associate a miner with a hardware to defend the sybil attack, as the hardware is costly to be replicated for gaining higher possibility of being elected. The PoS is running the same rule, it associates miner's money with the possibility of being elected.

Combining miner's identity with the hardware is an alternative approach to defend the sybil attacks as long as the hardware can prevent the identity from being forged. Trusted Computing seals the miner's identity inside the TPM, that makes each of the miner unique and unforgeable among the blockchain participants.

Unlike the consensuses without finality, e.g. PoW/PoS, who suffer from confirming time problem from several minutes to hours, Proof of Integrity uses round-robin or randomness as miner electing algorithm, because every miner has an unforgeable identity. This approach has a major advantage in that the block produced by a miner is finalized immediately, chain re-org is not required any more, long chain fork (consensuses in Bitcoin family) and uncle block (consensuses in Ethereum family) never happen. So, the blockchain state, e.g. account balance and smart contract result, can be retrieved and used at time when the latest block arrives.

Round-robin is simple and effective. Every miner in the miner list produces a block at a certain sequence. Let $m$ is a miner, and $M$ is all miners set in the list,

$$m_0, m_1, \ldots m_i, \ldots \in M$$

We define the order of list as $hash(m_i) < hash(m_{i+1})$.

At each round, a miner with index $i$ is active and produces a block, where $i$ is the index of a miner from the list, $h$ is the round number (block height), and $n$ is the total miner amount in the list.

$$i \leftarrow h \bmod n$$

Random is complex but more secure. The miner of next round is unpredictable until current round is determined. This makes DDoS targeted attacking is much difficult to be carried out, and reduces the opportunity of the miner's speculation. Verifiable Random Functions (VRFs) [45] is a good approach to generate a random number among all of the blockchain nodes, each of them can verify the random number independently. For any input seed $x$, $\text{VRF}sk(x)$ returns two values: a hash and a proof. The hash is a random value which is uniquely determined by $sk$ and $x$. The proof $\pi$ enables anyone who knows $pk$ to verify the *hash* is generated from $x$, without having to know $sk$.

$$x \leftarrow \text{hashOfBlock} (h - 1)$$
$$\langle hash, \pi \rangle \leftarrow \text{VRF}sk(x)$$
$$i \leftarrow hash \bmod n$$

### 3.2.4 Mining

Trusted Computing with Proof of Integrity protocol is an innovation to blockchain. The mining consensus is simplified that a miner shows its proof of integrity to other nodes and convinces them it is honest and not tampered. The detailed process is as following:
1) A miner is elected as an active miner for the next round according to *Miner Electing Method*.
2) The active miner produces a candidate block. This is similar to mining in other blockchain, including executing transactions, computing merkle root, generating block headers and packaging a final block, except that calculating difficulty is no longer required.
3) The active miner quotes its integrity values and generates Attestation Quote. The Trusted Computing TPM uses *quote* command to retrieve these values together with a signature, ensuring they are not forged or tampered.
4) The active miner composes the Attestation Quote and identity label into a final block, and then send the block out to all nodes in the blockchain network. Note that the identity certificate is no longer required since the certificate is known by other nodes (stored in the miner list) when *Miner Join* process completes.
5) When a node gets this block, it verifies if the block is eligible and correct. First, it checks if the integrity values are in the Integrity List. Second, it verifies if the integrity values signature is truly signed with the miner's identity. Last, it calculates the index $i$ according to *Miner Electing Method* to determine if this miner is eligible. When all of the verifications pass, the node accepts the block.
6) The integrity values and the signature (Attestation Quote) ensure the active miner

is honest and not tampered, subsequently, other node does not require to run the transactions any more. The active miner can put the transaction result into the block to allow Single Execution Model, as described in a later section.

Proof of Integrity constrains two aspects, the identity and the integrity. The identity has one to one relationship with a miner's hardware, and the integrity ensures the miner is honest and not tampered. The malicious miner, whose identity is not from a certified TPM or the integrity is not correct, will fail to join the blockchain network and cannot produce a valid block later.

### 3.2.5 Compatible Protocols

There would be more kinds of algorithm to elect the miner. [44] proposed RANDHOUND and RANDHERD for some certain degrees of distributed randomness used as node selection to generate block. A protocol used by a blockchain consensus is consistent in all nodes enforced by Proof of Integrity, malicious mining behaviors and tampered blocks from a dishonest node will be ignored and cannot break the security of the blockchain.

## 3.3 Enhance the Efficiency of Blockchain

### 3.3.1 Transaction

Trusted Computing with Proof of Integrity in blockchain can reduce the transactions execution effort significantly. Each node will trust the transaction result executed by the honest miner who has already checked the correctness, e.g. signatures, amounts and balances, etc. In most cases, the miner who is responsible to execute the transaction and calculate the result is called executing node (executing node is a miner node, it's used for representing execution purpose). The other node without executing transaction is called non-executing node.

Figure 2 shows transaction execution flow. The executing node produces a block containing the transactions and results, and then broadcast to the network. The other nodes can trust the transactions results by checking the miner's integrity.

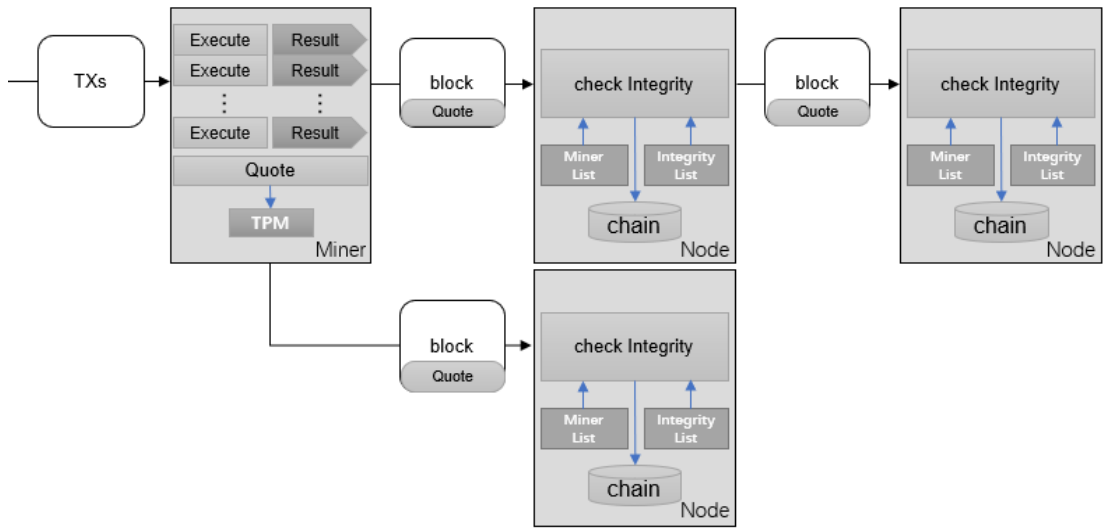

Figure 2: Transaction Execution

The efficiency of the transactions execution is higher, because the time complexity of executing transactions in a block is reduced from $O(n)$ to $O(1)$. The miner can produce a bigger block to increase the throughput without effort overhead in all nodes. Comparing to other blockchains who force transaction executions in every node, Truxen has outstanding effectiveness.

### 3.3.2 Smart Contract

Smart contract derives from the Bitcoin transaction script but is quite different in that it is Turing completed and is more powerful to do complicated business logic, as smart contract can be written in advanced language, like Golang, Java, Solidity, Serpent etc. Like transaction, executing smart contract is duplicated in each node, the result is calculated independently and then persisted into blockchain storage.

Trusted Computing with Proof of Integrity makes the smart contract execution simplified. A executing node executes contracts in the meantime it produces a block, and involve the results in the block, then broadcast them to the network. When a node receives the block, it checks the integrity, and then persisted the results into blockchain storage.

Only one executing node to execute the transactions and smart contracts in blockchain, is called **Single Execution Model (SEM)**.

### 3.3.3 Single Execution Model

Single Execution Model is a unique feature of Truxen. Comparing to other blockchains, it is advanced in the features of determinacy, efficiency and off-chain invocation.

In traditional smart contract, random number is not allowed because different nodes have different entropy sources or seeds, resulting in different result. However Single

Execution Model can safely generate random number within smart contract code that will not break the consensus. The random number can be used as nonce of signing, seed of a key, object identity and data grouping, etc. The Random Number Generator module in TPM is a best choice, as it has physical entropy source and is separated from the other parts of the system. The entropy sources are physically randomness, they could be the noises in the circuit, temperature of the chip, clock jitter and other type of physical events. The separation from other parts mitigates the security issues like seeds collision, comparison-based attack.

Single Execution Model has potential high performance and efficiency, and can increase the blockchain throughput. It distributes the workload among all the miners, while others save the computing resource and can do more jobs. Note that Single Execution Model requires the block involves result of executing transaction and smart contract, that increases the block size and networking effort by 10%-20% approximately, but decreases the computing load by 90%. In busy time, the block is enlarged to contain more transactions, but the computing load is barely up, as shown in Figure 3.

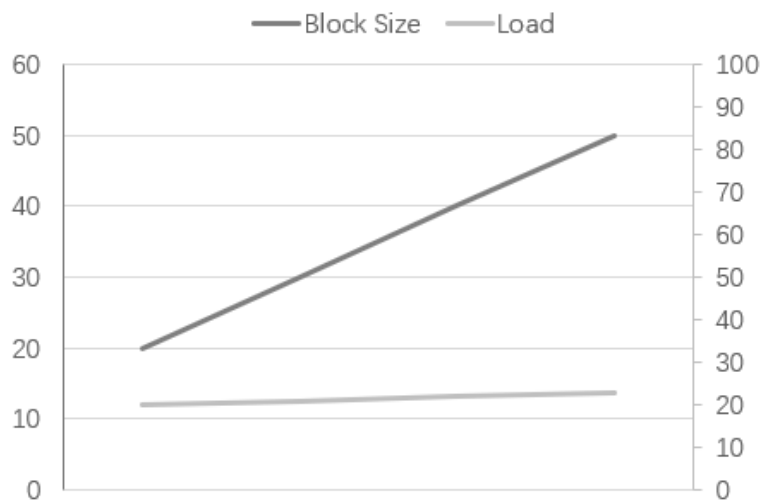

Figure 3: Block Size and Computing Load

Blockchain does not enable off-chain invocation natively, because the invocation runs from every node in the network, that is impossible for the off-chain system to accept such kind of remote calls. But off-chain invocation is essentially helpful to business logic to retrieve data or notify events. [29] proposed a method to invoke the website interface inside a smart contract through the trusted hardware, and the response is trusted and tamper-resisted. On the other way, Oracle as a workaround allows the event trigger mechanism to send remote calls to off-chain applications, and gets responses as a separated input of smart contract. Single Execution Model is a state-of-the-art approach to natively support off-chain invocation as it prevents the duplicated outbound requests and inconsistency responses. Transactions and smart contracts inherently run on an executing node which will initiate the remote call to off-chain system and get response synchronously, while other nodes do not do it.

## 3.4 Protection for Sensitive Data

The sensitive data, for example private keys, are critical to blockchain security, since they are the core ingredient of wallets or user identity. In most cases, the sensitive data are stored in file system which are prone to leak. Although the they could be protected by passcode, it is insufficient secure as the passcode could be weak and suffer from dictionary attack.

In Bitcoin, there is a kind of wallet which is a dedicated hardware and can generate and store private keys. The hardware wallet is the safest wallet because the private keys never leave the hardware. Similarly, in Trusted Computing, there is a Protected Location which can be used to store the sensitive data confidentially. The sensitive data is encrypted inside Trusted Platform Module and stored in its non-volatile memory. The encryption uses seed and key that never leave the platform, therefore there is no way to leak data.

The private key can be derived from Endorsement Key as seed, like deterministic wallet in Bitcoin, and it could be restored within the same hardware in case of lost. While other hardware cannot derive the same key as the seed is different. The alternative key generation is through Random Number Generator in TPM, and the key must be stored securely in Protected Location.

## 3.5 Network Topology

The nodes without Trusted Computing can collaborate in the blockchain network as well. So, there are two types of nodes, node with Trusted Computing and node without. Different nodes play different roles. Figure 4 shows the mixed nodes in the network.

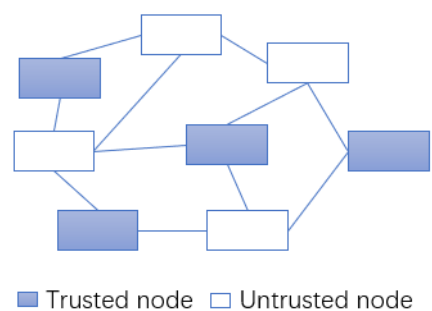

Figure 4: Mixed nodes in Truxen topology

The jobs processed by trusted nodes only are: generating blocks, quoting the integrity values, executing the transactions or smart contracts and calculating results. These actions need Trusted Computing to resist adversaries, Sybil attack, double spending, spamming transactions, malicious code, majority attack, etc. The consensus is initiated from trusted nodes, which play major role int blockchain to keep secure and consistent. The jobs processed by both types of nodes include sending transactions, holding blockchain storage, propagating the transactions and blocks, updating storage and checking integrity of the block. The untrusted nodes are usually used to relay transactions, inquire the transactions data and account balance.

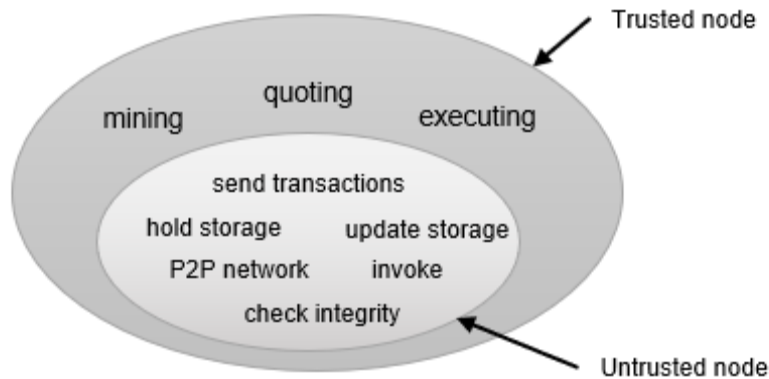

Figure 5: Relations between trusted node and untrusted node

## 4. RELATED WORK

Trusted Computing plays an important role in the context of blockchains. Several related papers and projects are presented for different purposes from data privacy, transaction security, and block producing algorithm. The domains they are trying to resolve reflects what the blockchain lacks of and what is evolving rapidly.

*Hyperledger Sawtooth* is using Intel SGX as a Trusted Execution Environment (TEE). It composes the block producing algorithm inside the TEE which is running on Intel CPU. *Sawtooth* uses a Proof of Elapsed Time (PoET) consensus that allows all miners of the blockchain to wait a randomized time at each of mining round, and the one whose waiting time is shortest wins the round, and then is eligible to produce a block. *Sawtooth* verifies the miner's legitimacy with Intel's IAS service and determines if a miner has been truly waited the amount of time it claims. The codes in Proof of Elapsed Time consensus must be separated from the other parts of blockchain as the consensus module is running inside TEE which uses a different development policies and tools.

*Proof of Luck* blockchain is built upon TEE. It consists of two functions PoLROUND and PoLMINE. At each of mining round, the node calls PoLROUND to get a round time. The shortest round time means a luckier number, and will enable the node to produce a block. Function PoLMINE prepares a block with previous block hash, and ensures the round time passes before the block is sent out. An attestation is sent to remote nodes at the meantime that allows the nodes to validate the correctness. A monotonic counter is setup in PoLMINE to prohibit concurrent invocations of the TEE.

*Ekiden* [46] is a confidentiality-preserving, trustworthy platform, it can offload smart contract execution from blockchain to TEE and perform data encryption. *Ekiden* can perform thousands of transactions per second due to the effective combination of blockchains and trusted hardware. *Ekiden* uses a so-called *compute nodes* as smart contract execution and attests the result on chain. Besides *compute nodes*, there are *consensus nodes* which maintain the blockchain ledger and do not require the TEE capability. *Ekiden* uses opensource RISC-V project Keystone [47] as the TEE runtime.

IBM Research proposed a TEE based hyperledger fabric solution [48]. They are using SGX enclaves to execute chaincode, maintain ledger and registry. Chaincode will be executed in *Chaincode Enclave* after remote attestation is performed and verified. And the state data are stored into ledge in a separated *Ledger Enclave*. The state data and chaincode operation can be encrypted using *PKcc* and decrypted using *SKcc* inside the enclave, which keeps the transaction and data secret from other peers.

*Proof of Authority* consensus is deployed in Ethereum Rinkeby test net. It assigns every miner an identity and chooses the miner to produce a block according to the identity, less computationally intensive than PoW approaches. When a new miner is joining the blockchain, existing miners will vote for it, major part of their decisions will determine the joining permission.

*Oracle* provides some abilities to interact with off-chain applications. *Oracle* is an off-chain agent that continually reads the events of blockchain transactions log and invokes the remote interface if certain conditions are met. *Oracle* usually is an asynchronized process, requesting and responding are in different threads, and it's the smart contract responsibility to check the response correctness and trustworthiness. *Town Crier* [49] uses TEE as an *Oracle* to retrieve the data feed for smart contract from outside system, which simplified the verification of remote response.

## 5. CONCLUSION

Trusted Computing enhanced blockchain simplified the consensus implementation, and it can increase the efficiency in transaction and smart contract execution, as well as makes blockchain more secure with encrypted protection. Additionally, it has the ability to invoke off-chain system, that makes blockchain more flexible and adoptable. Finally, it has in-deterministic capability that allows some degree of randomized logic in smart contract code.